\documentstyle[preprint,aps]{revtex}
\begin{document}
\title{Particle size effects in the antiferromagnetic spinel CoRh$_2$O$_4$}
\author{R. N. Bhowmik$^a$\footnote{e-mail:rnb@cmp.saha.ernet.in}, R.
Nagarajan$^{b}$ and R.
Ranganathan$^a$}
\address{$^a$Experimental Condensed Matter Physics Division,\\ 
Saha Institute of Nuclear Physics, 1/AF, Bidhannagar, Calcutta 700064, India\\
and\\ $^b$Tata Institute of Fundamental Research, Solid State Condensed Matter
Physics Division,Mumbai, India}
\maketitle
\begin{abstract}
We report the particle size dependent magnetic behaviour in the 
antiferromagnetic spinel CoRh$_2$O$_4$.
The nanoparticles were obtained by mechanical milling of bulk material,
prepared under sintering method.
The XRD spectra show that the samples are retaining the spinel structure.
The particle size decreases from 70 nm to 16 nm as the milling time increases
from 12 hours upto 60 hours.
The magnetic measurements suggest that the 
antiferromagnetic ordering at T$_N$ $\approx$ 27K exists in bulk as well as in 
nanoparticle samples. However, the magnitude of the magnetization below T$_N$
increases with decreasing particle size. Considering the fact that Rh$^{3+}$
has strong octahedral (B) site occupation and no change in T$_N$ of bulk and
nanoparticle samples, we believe that the observed magnetic enhancement is
not related to the cationic redistribution between
tetrahedral (A) and octahedral (B) sites of the spinel structure.
In our opinion, the observed effect is a consequence of decreasing
coherent length of antiferromagnetic coupled core spins and
increasing number of the frustrated shell in the core-shell model of
anoparticle. 
\end{abstract}
\section{Introduction}
The nanoparticle spinel ferrites are under the intensive
investigation in recent years because of their potential applications in
nanoscience and technology as high density magnetic recording media, magnetic
carriers in ferro fluids, magnetically guided drug carrier etc. \cite{Dian}. 
The theoretical interest for such type of materials are growing up to 
understand the structural and magnetic modifications taking place in a system
when the dimension of the particles (crystal size) are reduced to atomic scale
\cite{Tejada,Fine,Ganguli,Kodamascale}. Several novel phenomena like  
magnetic quantum tunneling \cite{Tejada}, superparamagnetism, surface spin 
canting \cite{Kodama}, grain boundary effect \cite{Bianco}, non-equilibrium 
cation distributions among the inequivalent lattice sites \cite{Hamdesh} 
are attracting the spinel ferrites. In spinel
lattices the anions (O$^{2-}$,S$^{2-}$ ions) form a cubic close packing, in
which the interstices are occupied by tetrahedral (form A sites or sublattice)
and octahedral (form B sites or sublattice) coordinated cations to the oxygen
anions. The competition between various type of superexchange 
interactions via O$^{-2}$ ions, {\it i.e.}, inter-sublattice superexchange 
interactions (J$_{AB}$) between ions of both sites and intra-sublattice 
superexchange interactions (J$_{AA}$ and J$_{BB}$) between ions of same site, 
shows a variety of magnetic states like ferrimagnet/ferromagnet, 
antiferromagnet, superparamagnet and spin glass in spinel oxides \cite{Fine}. 
When the particles size are reduced in the nanometer scale
a drastically different kind of magnetic behaviour were observed in spinel 
oxides in comparison with the bulk material \cite{Kodamascale,Finite}.  This 
has been explained in terms of site disorder,
{\it i.e.}, the cations redistribution between A and B sites
\cite{Sepel,Oliver} and finite size scaling effect \cite{Tang}.\\
It is established that for long range ferrimagnetic ordering in spinel oxide,
the necessary condition is ${\vert}J_{AB}{\vert} >> {\vert}J_{BB}{\vert} >>
{\vert}J_{AA} {\vert}$ \cite{Buschow}. 
However, if we compare the antiferromagnetic ordering temperature (T$_N$) of
two typical normal spinel antiferromagnets ZnFe$_2$O$_4$ with T$_N$ 
$\sim$ 10K (where only J$_{BB}$ exist) and CoRh$_2$O$_4$ with T$_N$ $\approx$
27K (where only J$_{AA}$ exist) \cite{Blass}, it can be understood that 
${\vert}J_{BB}{\vert} >> {\vert}J_{AA} {\vert}$ may be true only for long
range order ferrimagnetic spinels (where both A and B sites are 
occupied by magnetic ions) but not for all the cases, particularly, for the spinels with
magnetic ion only on A site. 
Unfortunately, most of the reports deal with the nanoparticle spinels where
either B or both A and B sites are occupied by magnetic ions
\cite{Hamdesh,Kodama,Sepel}.\\ 
It will be very interesting to investigate the particle size effect on
antiferromagnetic spinels with magnetic moment only on A sites.
Recently, M. Sato et al. \cite{Msato} reported the disappearance of
antiferromagnetic ordering at 33K of bulk Co$_3$O$_4$ spinel and the
appearance of a variety of magnetic phases like ferrimagnet, superparamagnet
and spin glass when the particle size is reduced to nano scale.
For Co$_3$O$_4$ spinel oxide, the B site is fully occupied by non-magnetic
Co$^{3+}$ (3d$^{6}$) ions and A site is occupied by magnetic Co$^{2+}$ ions, 
which gives rise to long range antiferromagnetic order due to 
Co$^{2+}$-O$^{2-}$-Co$^{2+}$ (J$_{AA}$) superexchange interactions with 
J$_{AB}$ = J$_{AA}$ = 0 \cite{Msato}. CoRh$_2$O$_4$ (structure:
(Co$^{2+}$)$_A$[Rh$^{3+}$]$_2$O$_4$) with T$_N$ $\approx$ 27K is an anlogus of 
Co$_3$O$_4$ (structure: (Co$^{2+}$)$_A$[Co$^{3+}$]$_2$O$_4$), where Co$^{3+}$ 
is replaced by non-magnetic Rh$^{3+}$ (4d$^6$) ions \cite{Blass}.\\ 
Recently, a significant research interest is focussing on the geometrically 
frustrated antiferromagnets (Ising or Heisenberg in nature).
The change of degeneracy and topology of the antiferromagnetic ground state
(Neel order) of such a system can show various kind of interesting magnetic 
properties such as quantum disordered ground states \cite{Petro}, quantum zero
-temperature spin fluctuation effect, where the
system do not order and remain in a "collective paramagnetic state" down to
zero temperature \cite{Villain}.
The degeneracy of the antiferromagnetic ground states can be reduced by 
introducing random non-magnetic dilution \cite{Henley} or by strain induced 
positional disorder \cite{Richert}. Even some authors introduced the concept
of ordering in geometrically frustrated system due to disorder while the
degeneracy is reduced in the system \cite{Petro,Henley}.
Mechanical milling is one of the most convenient method which can introduce
the positional disorder in the lattices and simultaneously reduce the particle
size of the material.\\
In this paper, we address the nature of magnetic order as a function of 
particle size in CoRh$_2$O$_4$ prepared by mechanical milling. The samples are
characterized by XRD and magnetization measurements have been performed using 
SQUID magnetometer.
\section{Experimental}
We have synthesized nano particles CoRh$_2$O$_4$ spinel oxide by mechanical
milling of the bulk material using Fritsch Planetary Mono Mill "Pulverisette 6".
For bulk material, the stoichiometric mixture of Co$_3$O$_4$(99.5\% from Fluka) 
and Rh$_2$O$_3$ (99.9 \% from Johnson Matthey) oxides was taken for
CoRh$_2$O$_4$ composition. The mixture was mechanically ground for 2
hours and was pelletized. The pelet was sintered at 1000$^0$C for 12
hours and at 1200$^{0}$C for 48 hours. The sample was then cooled to room
temperature at 2-3$^0$C/minute. A typical crystalline spinel structure was
confirmed by X ray diffraction (XRD) spectra using Philips 
PW1710 diffractometer with Cu K$_{\alpha}$ radiation. 
The bulk material was powdered using a 80 ml agate vial with 10 mm agate 
balls. We intentionally did not use the stainless bowl and balls to avoid any 
kind of contamination of transition metals (Fe, Cr, Ni). The samples were 
milled with ball to powder mass ratio 12:1 and at a rotational speed of 300
rpm. Small amount of samples were taken out from the bowl after 12 (mh12
sample), 24 (mh24 sample), 36 (mh36 sample), 
48 (mh48 sample) and 60 (mh60 sample) hours of milling for our studies. The dc magnetization measurements 
were performed using SQUID (quantum design) magnetometer.
\section{Results and Discussion}
\subsection{Structural properties}
The X ray diffraction spectra (Fig.1) show that the crystalline nature of bulk
CoRh$_2$O$_4$ decreases with increasing milling time. 
It should be noted that the crystalline peaks of milled samples, as shown for 
311 line (Fig. 1b), shows small shift to higher scattering angle (2$\theta$) 
with respect to the bulk sample. However, there is no extra  
lines in XRD spectra for as milled samples in comparison with bulk sample.
This suggests that small amount of lattice disorder or lattice strain is 
introduced in the system as the particle size is reduced by mechanical milling
but without changing the crystal symmetry of spinel structure
\cite{rnbjms,Cili}.
The decrease of lattice parameter (see Table 1) suggest that it is related
with the decrease of particle size as a function of milling time \cite{Naran}. 
The transmission electron micrographs (TEM) confirm the
decrease of particle size from 70 nm (12 hours milling) to 16 nm (60 hours
milling) (see Table 1). The broad lines in XRD spectra is due to this decrease 
of particle size, where the thermal fluctuation of the small coherent 
(crystalline) zones broaden the peaks \cite{Cili}. 
The XRD peak shift of the milled samples suggest that the non-uniform 
microstrain developed at the lattice sites during mechanical process may
also contribute such type of peak broadening \cite{Bianco}.
In literature the small shift of XRD peak is very often neglected. But a 
critical observation of this type of shift is very important in correlating 
the physical properties with mechanical strain induced change in a sample
\cite{rnbjms}.
\subsection{Magnetic properties}
The inset of Fig. 2 (left scale) shows the zero field cooled (ZFC) and field 
cooled (FC) magnetization data for CoRh$_2$O$_4$ bulk sample, 
measured at 100 Oe dc magnetic field. The bulk sample shows antiferromagnetic 
ordering at T$_{N}$ $\approx$ 27.5K$\pm$0.5K
and magnetic irreversibility between ZFC and FC magnetization below T$_N$. The 
ZFC magnetization data at T$> 50K$ are well fitted with Curie-Weiss law (Fig.2 
inset, right scale)
\begin{equation}
\chi = \frac{C}{T-\theta_{A}}
\end{equation}
The Curie constant (C = N$\mu_{eff}^{2}$/3k, N is the number of  
CoRh$_2$O$_4$ formula unit per gram of the sample) gives the effective 
magnetic moment ($\mu_{eff}$) = 4.60$\pm$0.10 $\mu_{B}$) per formula unit for 
the bulk sample.
The asymptotic Curie temperature ($\theta_A$) is $\approx$ -(45$\pm$ 2)K. 
These values are consistent with the reported values $\mu_{eff}$ = 
4.55 $\mu_{B}$) and -30K for bulk CoRh$_2$O$_4$ spinel \cite{Blass}. 
The negative value of $\theta_{A}$ indicate that on lowering the temperature
the antiferromagnetic ordering is saturated at T$_N$ $\approx$ 27.5K and the
system shows strong antiferromagnetic ordering below 27.5K. 
This indicate the magnetic phase of our bulk CoRh$_2$O$_4$ sample is
consistent with the reported one. 
Interestingly, all the milled samples (with smallest particle size $\sim$ 16
nm) are showing (Fig.2, log-log scale) antiferromagnetic ordering 
at T$_N$ $\approx$ (27.5$\pm$0.5)K 
with magnetic irreversibility between ZFC and FC magnetization when
temperature decreases below T$_N$. The larger value of FC magnetization than 
ZFC magnetization below T$_N$ 
suggests the field induced metastable magnetic state during field cooling 
process of the samples \cite{Matt}. It is also found that the ZFC
magnetization data at T $>$ 50K fit with Curie-Weiss law (Fig.2 inset, right 
scale), as an example shown for mh60 sample, for all the milled samples. The 
effective magnetic moment ($\mu_{eff}$) and $\theta_{A}$ values are shown in 
Table 1. We see that the effective magnetic moment value is increasing
with decreasing the particle size. Similar kind of magnetic enhancement was
observed by F. Liu et at. \cite{Khan} and was attributed as a function of the
reduction of coordination number of the surface spins when the dimensionality 
of feromagnetic particles were reduced. 
The ratio of $\theta_A$
and T$_N$ is always greater than 1. According to ref. \cite{Palm}, if this
ratio quantify the degree of magnetic frustration in a geometrically 
frustrated antiferromagnet, then we can say that geometrical frustration and
the instability of antiferromagnetic ordering is
increasing with decreasing the particle size by mechanical milling of bulk
CoRh$_2$O$_4$ spinel. The main characteristic feature is that both the ZFC and
FC magnetization are increasing at T$ << T_{N}$, which is very similar to 
paramagnetic or superparamagnetic \cite{Fiorani} or ferrimagnetic contribution
\cite{Msato} in the samples. Even, the increase in the magnitude of 
magnetization below T$_N$ can be assigned due to the increasing number of 
uncompensated/frustrated spins \cite{Sera} as the particle size decreases. 
However, the temperature dependence of inverse of zero field cooled 
susceptibility (H/M$_{ZFC}$) at T $<T_N$ shows downward curvature in Fig.3. 
Interestingly, inset of Fig.3 shows that H/M$_{ZFC}$ $\propto$ T$^{\alpha}$ 
below 10K and the constant value $\alpha$ increases with decreasing particle 
size. This indicates that the magnetization below
T$_N$ are something different from a typical paramagnet or superparamagnetic
behaviour, where inverse of susceptibility should be linear with temperature
and $\alpha$ should be 1. 
M. Sato et al. \cite{Msato} suggested similar kind of magnetic behaviour
below T$_N$ due to the appearance of ferrimagnetic phase when the particle 
size of the antiferromagnetic spinel Co$_3$O$_4$ was reduced to 15 nm.\\
The most important feature in Fig.4 is that the excess amount of zero field
cooled magnetization of milled samples $\Delta$M$_{mb}$ (=
M$^{milled}_{zfc}$-M$^{bulk}_{zfc}$) over the bulk
sample not only increases at T $<$ T$_N$, but also depend on the particle size. 
We have found in Fig.4 inset that $\Delta$M$_{mb}$ vs T follows a scaling law 
at T$<<T_{N}$ in the form
\begin{equation} 
{\Delta}M_{mb} = ({\Delta}M_{mb})_{0}T^{-(0.937{\pm}0.002)}
\end{equation}
where the constant ($\Delta$M$_{mb}$)$_0$ depends on the particle size and
linearly increases
as 8.952$\times$10$^{-3}$, 1.475$\times$10$^{-2}$, 2.111$\times$10$^{-2}$,
2.821$\times$10$^{-2}$ and 3.576$\times$10$^{-2}$ (in emu/g unit) for mh12,
mh24, mh36, mh48 and mh60 samples, respectively.\\
The excess in of FC magnetization over the ZFC magnetization (Fig.5a) 
,{\it i.e.}, $\Delta$M$_{FZ}$ = M$_{FC}$-M$_{ZFC}$ increases below T$_N$
in a typical manner which has similar character as the uncompensated 
interfacial antiferromagnetic spins exhibit in Ni$_{81}$Fe$_{19}$/CoO
bilayers \cite{Takano}. In case of spinel oxides, the surface cations have 
various number of next nearest neighbours on both A and sites. When the 
particle size are reduced to nanometer range, some of the exchange bonds are 
broken and coordination number to oxygen ions are also decreased. This results 
a distribution of net exchange fields, which controll the
surface magnetism of the particle \cite{Kodama}.
This exchange field is proportional to 
the spin density of the uncompensated antiferromagnetic spins at the surface 
\cite{Takano}. The magnitude of $\Delta$M$_{FZ}$ will depend on the number of
uncompensated spins and also
the exchange interactions between field aligned (uncompensated surface) spins 
and the antiferromagnetic core spins. 
This argument invokes the core/shell picture \cite{Kodama} for our samples,
where the shell thickness, consisting of uncompensated spins, is increasing 
with decreasing the particle size by decreasing the core volume.\\
The zero field cooled magnetization at 100 Oe, 1 kOe and 1 Tesla field  
for rh48 sample (particle size $\sim$ 19 nm) (Fig.5b) do not show any 
appreciable change of T$_N$ with fields. This suggests
that dominant antiferromagnetic order (LRAO) still exists for the nano 
particle samples. However, it is the fact that long range antiferromagnetic
ordering is proportional to the divergence of magnetization at T$_N$.
Qualitatively, we can say, LRAO is proportional to the difference between peak 
magnetization (M$^{peak}_{ZFC}$) at T$_N$ and minimum of magnetization 
(M$^{min}_{ZFC}$) below T$_N$. Following this argument, we see (Fig.5a, inset)
the difference 
between peak magnetization and minimum magnetization below T$_N$, i.e., 
${\Delta}M_{pm}$ = (M$^{peak}_{ZFC}$-M$^{min}_{ZFC}$)/M$^{peak}_{ZFC}$, 
reduces from 27\% (for bulk sample) to 0.5\% (for mh60 sample). This confirm 
that although antiferromagnetic ordering is still observed below T$_N$, the 
magnetic disorder is increasing when the particle size decreases 
by mechanical milling \cite{Richert}.\\ 
Fig.6 shows the zero field cooled magnetization as a function of magnetic
field at 5K for all the samples. The straight line nature of M vs H plot for
H = -3T to +7T range shows a typical antiferromagnetic bulk sample. The
antiferromagnetic nature is still very prominent for mh12 sample. But the
down ward curvature of the M vs H curve (see for mh36 and mh60 samples) in 
the positive field range suggests that some magnetic contribution is 
increasing as the particle size decreases. From the Arrot plot (M$^{2}$ vs
H/M),we have found no spontaneous magnetization for any samples, where as the 
linear extrapolation of the data (for H$\geq$ 4 Tesla) to the M axis gives 
some finite values of M$_{0T}$ for all milled samples. The M(H) data,
therefore, confirm that there is no ferromagnetic ordering in system. The
magnetic contribution arising in decreasing the particle size can be
attributed as disorder and dilution effect in antiferromagnetic spinel
\cite{Msato}. The increase of M$_{0T}$ (Fig.6 inset) with increasing the 
milling time indicates that although the samples does not 
show any ferromagnetic spontaneous magnetization but field induced magnetic
ordering is possible for antiferromagnetic nano particles \cite{Zhito}.
Fig.7 shows the M vs H data at different temperatures for the 48 hours milled 
sample. The linear extrapolation of M for H$\geq$ 4T to H= 0T axis shows 
(Fig. 7 inset) that the magnetization (M$_{0T}$) first decreases with increasing
temperature down to $\approx$ 16K and then increases upto 27K. The temperature 
dependence of M$_{0T}$ is very similar to the temperature dependence of
magnetization at T$< T_{N} {\approx}$ 27.5K for the same sample. This type of
magnetic behaviour suggests that there is certainly a competition between
antiferromagnetic order and magnetic order due to disorder effect in the nano 
particle samples \cite{Henley}. Further, it can be suggested that the disorder
effect will dominate as the temperature is well below of T$_N$.
\section{Summary}
The bulk CoRh$_2$O$_4$ spinel is an antiferromagnet with ordering temperature 
T$_N$ $\approx$ 27.5K. When the particles size of CoRh$_2$O$_4$ are reduced by
mechanical milling, the signature of antiferromagnetic order at $\approx$ 
27.5K are still observed upto particle size $\approx$ 16 nm. In case of
nano particles, the magnetization below T$_N$ is enhanced with respect to the
bulk sample, which is followed by a scaling law. 
Since the antiferromagnetic ordering temperature at T$_N$
is unchanged, the enhancement in magnetization can not be attributed due to
the site exchange \cite{Braber} between Co$^{2+}$ and Rh$^{3+}$ ions when 
the particle size decreases by mechanical milling. Under this circumstances,
the tetrahedral (A) sites occupy magnetic Co$^{2+}$ ions and octahedral (B) 
sites occupy non-magnetic Rh$^{3+}$ ions and excludes the possibility of
convensional ferrimagnetic contribution in this system. The temperature
dependence of the inverse suseptibility below 10K also suggest that the 
enhancement of magnetization is not due to typical superparamagnetic 
contribution of the nano particles.\\ 
Therefore, we are introducing the core-shell structure of the nano particles
\cite{Kodama}. The core consists of 
antiferromagnetic spins and shell consists of few layers of surface spins.
The surface spins are coupled by superexchange interactions (via O$^{2-}$ 
ions) to the core spins. In case of bulk sample the length scale of
antiferromagnetic interactions can span upto many particles.
When the particle size is reduced to nanometer scale by mechanical milling,
some of the A-O-A superexchange bonds are broken and become frustrated. These
frustrated bonds (surface spins) will create exchange anisotropy field at the
interfacial surface \cite{Kodama,Takano}. This type of anisotropy field will 
give rise a preferential magnetic ordering of the loosely bound shell spins, 
whereas the tightly bound core spins will remained as antiferromagnetically
aligned.\\
\section{Conclusions}
Based on our dc magnetic measurements, it can be concluded that the total
magnetization of the nanoparticle M = M$_{core}$ +M$_{shell}$, where 
M$_{core}$ is
magnetic contribution from core spins and M$_{shell}$ is magnetic contribution
from shell spins. The competition between magnetic ordering of shell spins and 
the antiferromagnetic ordering of core spins guide the magnetic behaviour of 
our samples. The shell thickness is increasing in-expense of core volume when 
the particle size decreases. This is related to the decrease of coordination 
number of the surface (shell) spins and increase of magnetized state of the
surface spins due to increasing random exchange fields, as the size of the 
particle decreases. Consequently, the magnetic momemt will be enhanced in
nano particles. This is called disorder induced magnetic ordering in 
antiferromagnetic nano particle. 
\vskip 0.3 cm
\noindent Acknowledgement:
One of the authors RNB thanks Council of Scientific and Industrial Research 
(CSIR, India) for providing fellowship [F.No.9/489(30)/98-EMR-I ]. 
\newpage
Table 1. Particle size (from TEM photographs), Lattice parameter ($\AA$) (from XRD data), 311 peak 
position (from XRD data), effective magnetic moment ($\mu_{eff}$) (from M vs T
data), Asymptotic Curie temperature ($\theta_{A}$) (from M vs T data) as
a function of milling hours.\\
\begin{tabular}{c c c c c c c}
\hline\hline
sample & milling time & particle size & a($\AA$)& 2$\theta$ (deg) &
$\mu_{eff}$ ($\mu_B$ unit) & $\theta_A$ (K)\\\hline
bulk & 0h & few $\mu$m &8.465$\pm$0.002&35.47 & 4.599 & -44.23 \\
mh12 & 12h & 70 $\pm$1 nm &8.485$\pm$0.002&35.56 & 4.603 & -42.80 \\
mh24 & 24h & 50 $\pm$1 nm &8.427$\pm$0.002&35.75 & 4.609 & -42.05 \\
mh36 & 36h & 32 $\pm$1 nm &8.449$\pm$0.002&35.71 & 4.627 & -41.84 \\
mh48 & 48h & 19 $\pm$1 nm &8.468$\pm$0.002&35.67 & 4.653 & -43.81 \\
mh60 & 60h & 16 $\pm$1 nm &8.459$\pm$0.002&35.64 & 4.755 & -51.00 \\
\hline\hline
\end{tabular}

\end{document}